\documentclass[conference,compsoc]{IEEEtran}
\usepackage{subcaption}
\usepackage{graphicx}
\usepackage{amsmath,amssymb}	\interdisplaylinepenalty=2500
\usepackage{algpseudocode}

\ifCLASSOPTIONcompsoc
  \usepackage[nocompress]{cite}
\else
  \usepackage{cite}
\fi
\ifCLASSINFOpdf
\else
\fi
\begin{document}
%
\title{Human Uncertainty and Ranking Error\\[.5ex]  \Large{- The Secret of Successful Evaluation in Predictive Data Mining -}}

\author{\IEEEauthorblockN{Kevin Jasberg}
\IEEEauthorblockA{Institute of Information Science\\ University of Duesseldorf, Germany\\Email: kevin.jasberg@uni-duesseldorf.de}
\and
\IEEEauthorblockN{Sergej Sizov}
\IEEEauthorblockA{Institute of Information Science\\ University of Duesseldorf, Germany\\Email: sizov@hhu.de}
}



%


\maketitle

\begin{abstract}
One of the most crucial issues in data mining is to model human behaviour in order to provide personalisation, adaptation and recommendation. This usually involves implicit or explicit knowledge, either by observing user interactions, or by asking users directly. But these sources of information are always subject to the volatility of human decisions, making utilised data uncertain to a particular extent.

In this contribution, we elaborate on the impact of this human uncertainty when it comes to comparative assessments of different data mining approaches. In particular, we reveal two problems: (1) biasing effects on various metrics of model-based prediction and (2) the propagation of uncertainty and its thus induced error probabilities for algorithm rankings. For this purpose, we introduce a probabilistic view and prove the existence of those problems mathematically, as well as provide possible solution strategies.

We exemplify our theory mainly in the context of recommender systems along with the metric RMSE as a prominent example of precision quality measures. 
\end{abstract}


%
\IEEEpeerreviewmaketitle


\section{Introduction}

Data mining is an integral part of our modern information society.
Over the last decade, a multitude of algorithms and approaches have been developed in this thriving field of research in order to improve the modelling and prediction of human behaviour. These efforts are motivated by many practical applications, including various recommender systems, content personalisation, targeted advertising, along with many others. For all these systems, the human being is itself the main source of information, while the necessary information is obtained either implicitly by observing user interactions or explicitly by questioning a user.

\begin{figure}[t]
    \centering
        \includegraphics[width=\linewidth]{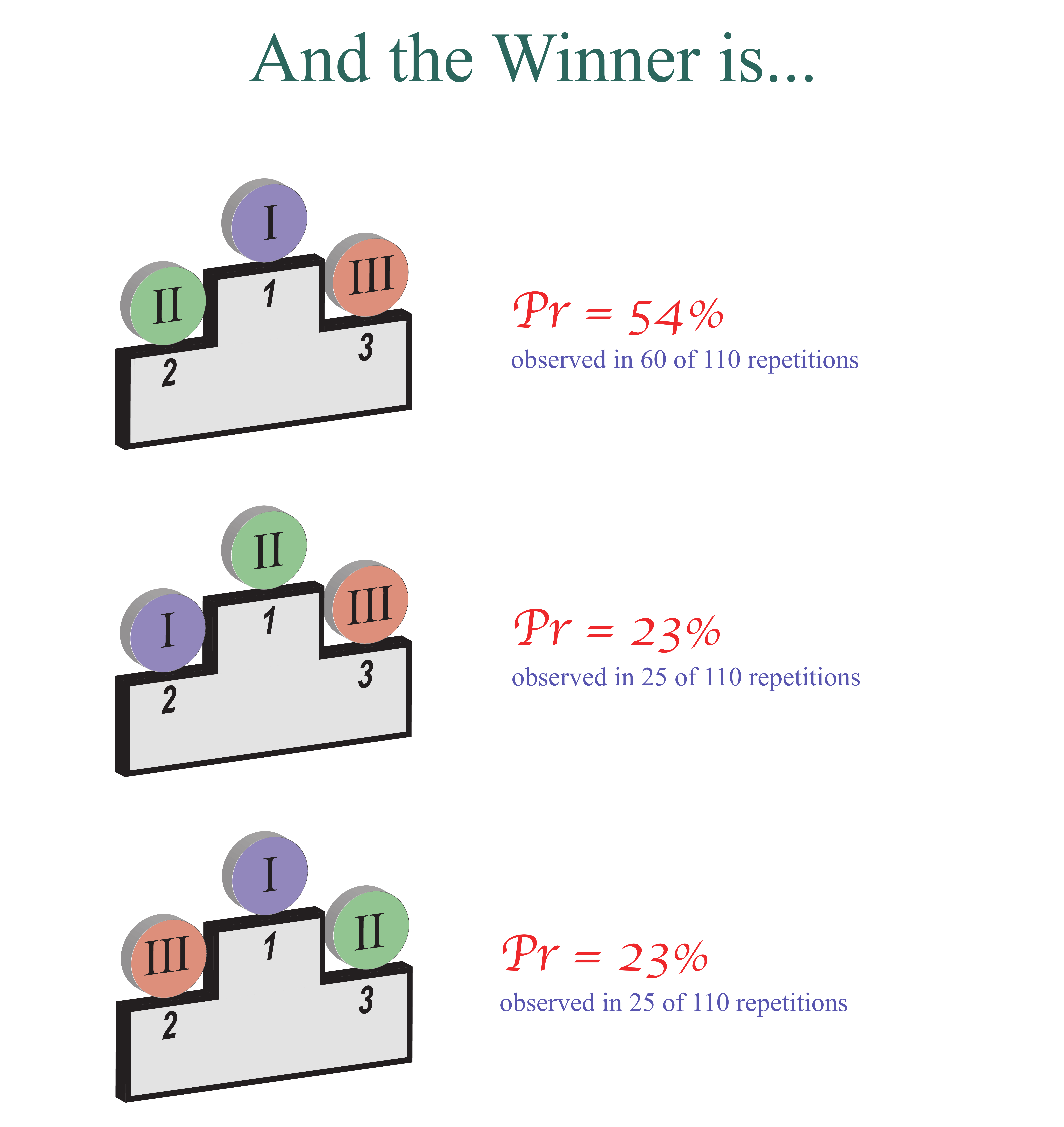}
    \caption{\small Rankings of Algorithms can vary with re-evaluation since users provide volatile feedback when asked again.}
             \label{fig:EyeCatcher}
\end{figure}

We take this as an occasion to ask how people really make decisions and how they communicate them while interacting with information systems. It is commonly known that human decisions are not constant, but subject to certain fluctuations, depending on the particular situation, mood, media literacy, and other biasing factors such as the interface. This quite human feature is also found within the databases of our information systems, especially in the case of explicit evaluations of sales products or new system developments. It has recently been shown for example, that users provide inconsistent ratings when requested to rate same films at different times \cite{Hill}. This \textbf{human uncertainty}, as we understand it in this contribution, appears to be a characteristic feature of the cognitive process of decision making \cite{Friston} which influences its outcome, making it circumstantial and temporally unstable - the outcome appears to be more or less fluctuating randomly when repeating a decision making. Consequently, we may assume that observed decisions are drawn from individual distributions \cite{FristonNature, Friston, delia}.

However, it is precisely this circumstance that makes the comparative assessment of data mining approaches - that you and I develop - even more difficult than it already is. The disillusioning question which arises from this reflections is whether a measured difference in a particular prediction quality metric is indeed due to a difference in a system's quality or merely an artefact of the unavoidable human uncertainty.
We will show that it may be possible to even reverse rankings by repeated measurement. This basically means, that a particular system may achieve a better metric score than another one in the first trial, but a worse score in the case of repetition. The logical consequence of this observation is that, unfortunately, we must part from the idea of accurate and reliable evaluations of data mining approaches. The tangible elaboration of this issue and its implications is the key concern of this contribution.\\[.5ex]

\begin{figure}[t]
    \centering
        \includegraphics[width=\linewidth]{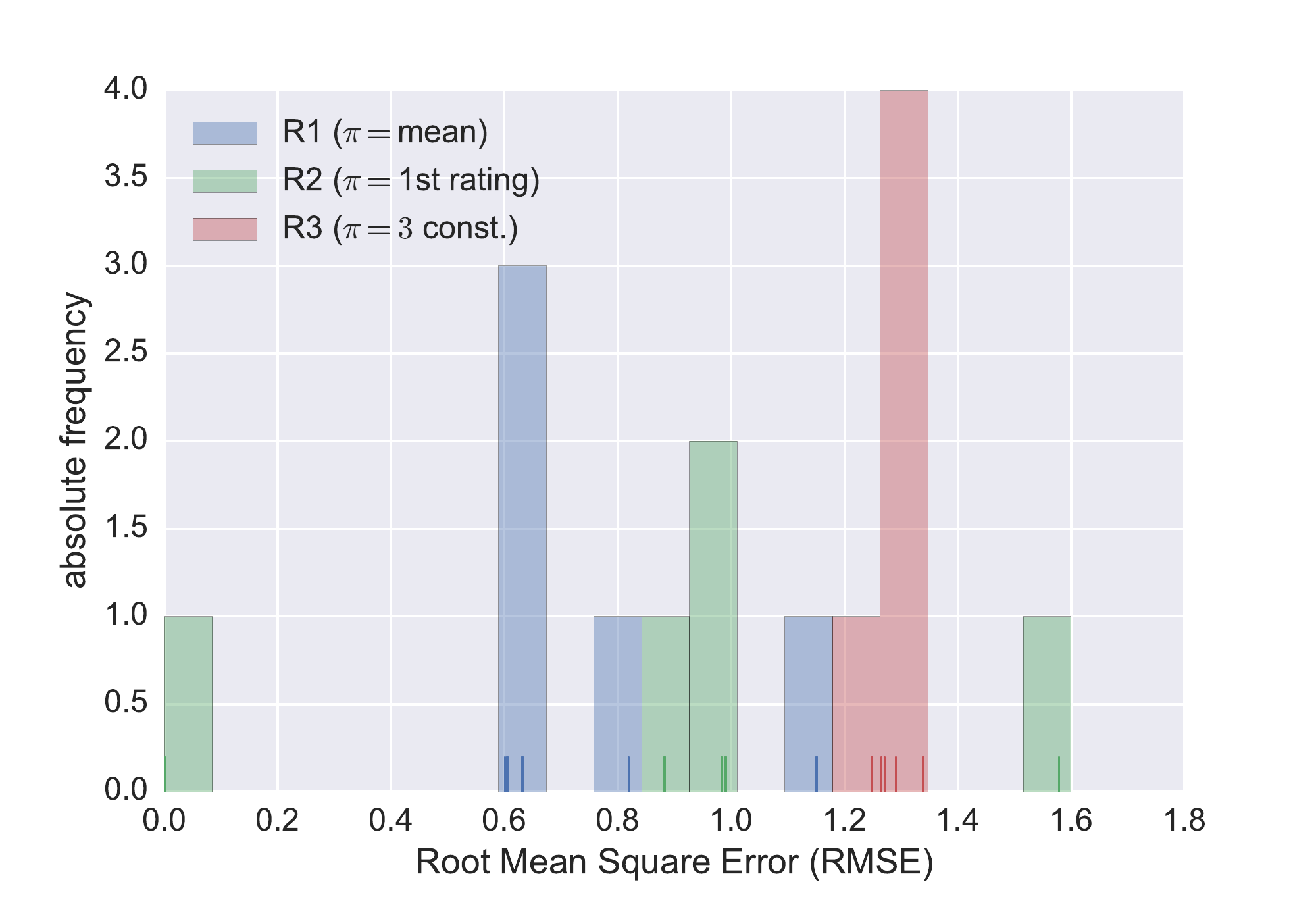}
    \caption{\small Uncertain user ratings induce volatile RMSE scores (user study).}
             \label{fig:Intro}
\end{figure}

\textsc{Motivating Example.} In a systematic experiment with real users, individuals rated theatrical trailers multiple times. It turns out that only 35\% of all users show constant rating behaviour, whereas about 50\% use two different answer categories and 15\% of all users make use of three or more categories. Based on these observations, we compute the Root Mean Square Error (RMSE) as a widely used metric for prediction quality for three different recommender systems, defined by their predictors
\begin{eqnarray*}
\text{Recommender R1} & & \pi:=\text{\small mean of ratings from user to item} \\
\text{Recommender R2} & & \pi:=\text{\small 1st ratings from user to item} \\
\text{Recommender R3} & & \pi:=\text{\small 3 const.}
\end{eqnarray*}
and for each rating trial. Figure \ref{fig:Intro} depicts those RMSE outcomes and their frequency in an absolute histogram. It becomes apparent at once that the RMSE itself yields a particular degree of uncertainty, emerged from uncertain user feedback. When ranking these recommender systems, Figure \ref{fig:Intro} allows for three possible results that emerge with different frequencies. The problem is most obvious for recommender R2 (green) as it could be both, the best or the worst recommender, although it operates for the same users rating the same items. Thus, the question for a comparison drastically changes, i.e. we are no longer looking for the only ranking available, but have to ask ourselves how likely any of these possible rankings actually is. 
This issue is addressed by Figure \ref{fig:EyeCatcher}, which depicts the possible rankings from our experiment combined with their frequency of occurrence. It turns out that the probability for each of these rankings is not negligible and no matter what ranking we finally opt for, there is always a certain chance of error for this decision.\\[.5ex]

%

\textsc{The Problem.} The matter of human uncertainty - if not explicitly considered - is that any improvement to an existing system or even the assessment of different systems might not be statistically sound. 
This, in particular, has financial implications when money is invested in
\begin{enumerate}
\item a supposedly better system whose superiority is just due to uncertainty.
\item the further development of an existing system when results are merely an overfitting. 
\end{enumerate}
This naturally evinces the necessity of a well-developed theory of distinguishing algorithms - given a particular metric for evaluation - when human uncertainty is involved.\\[.5ex]

\textsc{Our Objective.} 
The goal of this article is a profound elaboration of distinguishability in comparative assessments within the sub-field of human centred predictive data mining. In addition, we would like to find possible solutions and new ways of thinking to support differentiated comparisons between different approaches, which will make evaluations statistically more sound.\\[2ex]


\section{Related Work}

\textsc{Information Systems and Assessment.}
The central role of information systems led to a lot of research and produced a variety of techniques and approaches\cite{InfoSys}. Here, we focus especially on recommender systems which are comprehensively described in \cite{Jannach, Handbook}. For the comparative assessment, different metrics are used to determine the prediction accuracy, such as the root mean squared error (RMSE), the mean absolute error (MAE), the mean average precision (MAP) along with many others \cite{Herlocker, Bobadilla, workshop12}. These accuracy metrics are often criticised\cite{McNee} and various researchers suggest that human computer interaction should be taken more into account\cite{McNee2,Knijnenburg}. 
With our contribution, we extend existing criticism by an additional aspect that has little discussed so far.
Although we exemplify our methodology in accordance with the RMSE, the main results of this contribution can be easily adopted for alternative assessment metrics without substantial loss of generality, insofar they require for (uncertain) human input.\\[.25ex]

\textsc{Dealing with Uncertainties.}
The relevance of our contribution arises from the fact that the unavoidable human uncertainty sometimes has a vast influence on the evaluation of different prediction algorithms \cite{LikeLikeNot, noise2}.  The idea of uncertainty is not only related to predictive data mining but also to measuring sciences such as metrology. Recently, a paradigm shift was initiated on the basis of a so far incomplete theory of error \cite{Grabe, Buffler}. In consequence, measured properties are currently modelled by probability density functions and quantities calculated therefrom are then assigned a distribution by means of a convolution of their argument densities. This model is described in \cite{GUM}. A feasible framework for computing these convolutions via Monte-Carlo-Simulation is given by \cite{GUMsupp1}. We take this as a basis for our own modelling of uncertainty for addressing similar issues in the field of computer science. To derive a pragmatic and easy to handle theory, we will refer to the Gaussian Error Propagation which is commonly used in physics as well \cite{Ku,Bevington,Taylor}.\\[.25ex]

\textsc{The Idea of Human Uncertainty.} 
Probabilistic modelling of human cognition processes is quite common to
the field of computational neuroscience. In particular, aspects of human
decision-making can be stated as problems of probabilistic inference \cite{FristonNature}
(often referred to as ``Bayesian Brain'' paradigm). Besides external
influential factors, the belief precision is influenced by biological
factors like current activity of dopamin cells \cite{Friston}. In other words, human
decisions can be seen as uncertain quantities by nature of the underlying
cognition mechanisms. Recently, this idea has been adopted for various
probabilistic approaches of neural coding \cite{BayesianBrain}.
In parallel, many methods of predictive data mining employ probabilistic
(e.g. Bayesian) models for approximating mechanisms of human decisions
based on prior observations as training data. At the same time, common
evaluation approaches still use non-random quality metrics and thus do not
account for possible decision / ranking errors in a natural way. As a
consequence, we systematically tackle both, observed user responses and
resulting quality of the evaluated predictor as random quantities. This
allows us to elaborate on the impact of human uncertainty and provide solutions
for a more differentiated and objective assessment of predictive models.\\[.25ex]

\textsc{Experimental Designs.}
The complexity of human perception and cognition can be addressed by means of latent distributions \cite{delia}. This idea is widely used in cognitive science and in statistical modelling of ordinal data \cite{cub}. 
We adopt the idea of modelling user uncertainty by means of individual Gaussians following the argumentation in \cite{GaussModel} for constructing our individual response models. The methodology applied in our experiments is adopted from experimental psychology \cite{psycho} and works on repeating rating scenarios for same users-items-pairs as done before in \cite{RateAgain}.

\section{Modelling human uncertainty}
In this section, we embed human uncertainty into a mathematical construct and introduce an approach for determining the propagation of human uncertainty for a given evaluation metric. Therefore, we first develop a general framework which will then be illustrated for the RMSE as a prominent example for such evaluation metrics.

\subsection{Changing Paradigms}
As mentioned above, various experiments \cite{RateAgain,Hill} along with our own have shown that users are scattering around their true value of preference. Consequently, we may assume that observed decisions are drawn from individual distributions, as a result of complex cognition processes, and influenced by multiple factors (e.g. mood, media literacy, etc.) \cite{delia}. Therefrom, a paradigm shift has to be carried out, which is similar to the recent change of perspectives on measurement errors in metrology \cite{Buffler}: Every measurable quantity that is somehow related to human cognition is no longer considered as a single score (point-paradigm) but rather as a whole interval of possible values (set-paradigm) that is somehow distributed (distribution-paradigm). In the context of this paper, we will, therefore, consider user ratings as random variables. On this basis, we develop statistical methodologies that are to be explored hereinafter.  

\subsection{Composed Quantities}
Composed quantities are quantities that compute as a continuous function $Z=g(X_1,\ldots,X_N)$
of $N$ of uncertain arguments $X_\nu$ (random variables) and $Z$ hence becomes a random variable itself. This reasoning can be understood heuristically: A single random variable $X_\nu$ can take a variety of possible values $x_\nu$. 
Now having $N$ random variables, there is a plenty of possible values for $(x_1,\ldots,x_N)$ where each of these possibilities specify one value of $Z$  by means of $z = g(x_1,\ldots,x_N)$. Thus, the distribution of $Z$ emerges as a convolution of $N$ underlying density functions with respect to the mapping $g$ \cite{GUM, GUMsupp1}.

To be more specific: The well-established precision metrics MAE, MSE as well as the RMSE are prominent examples of such composed quantities. In order to exclude possible misjudgements during assessment, it is mandatory to determine the density that emerges for such a metric, i.e. we have to compute the convolution of all underlying feedback distributions.
As demonstrated in \cite{Chan}, exact solutions for RMSE-like quantities are quite laborious to find and difficult to implement. To overcome this problem, two standard techniques of experimental physics and metrology can be used, which are briefly presented below:\\[.2ex]

\textsc{Monte-Carlo-Simulation.} Using a Monte-Carlo-Simulation, we consider a vector of random variables $(X_1,\ldots,X_N)$ and simply compute $\tau$ pseudo-random-outcomes $(x_1,\ldots,x_N)$. For each outcome we are thus able to calculate $z = g(x_1,\ldots,x_N)$ numerically. The set $\{z\}$ can therefore be seen as the numerical representation for the distribution of $Z$. We obtain its density function by applying the Maximum-Likelihood-Method for appropriate distribution families. However, the disadvantage of this approach is that we are facing a blatant run-time problem as soon as we are entering the realm of big data. To compute the RMSE for a simple recommender system on the Netflix test record ($N = 2.8\cdot 10^6$ ratings), we needed about 35 hours using a state-of-the-art computing node. Having this in mind, Monte-Carlo-Simulation can unfortunately not be deemed to be rather feasible, although it produces excellent results. So, we will use this method only for confirming our mathematical approximations. \\[.2ex]

\textsc{Gaussian Error Propagation.} 
The advantage of big data is that it legitimises the use of the central limit theorem as long as we choose distributions with finite variances as the underlying data model, i.e. the type distribution of each $X_\nu$ (e.g. Gaussian, beta, etc.).
Since big data and the central limit theorem are supporting the assumption of normality of the resulting convolution, we simply need to approximate the expected value and the variance of a given composed quantity. Having such a quantity $Z=g(X)$ with $g\in C^\infty(\mathbb{R})$, the core of our estimation is to expand $g$ into its Taylor series. Due to the linearity of the expected value, we yield
\begin{eqnarray} \label{eq:TaylorExpect}
\mathbb{E}[g(X)]  
&=&  \mathbb{E} \left[ \sum_{k=0}^\infty \frac {g^{(k)}(\mu)}{k!} \, (X-\mu)^k  \right]  \nonumber \\
&=&  \sum_{k=0}^\infty \frac {g^{(k)}(\mu)}{k!} \mathbb{E}\left[(X-\mu)^k\right]  \nonumber \\
&=&  \sum_{k=0}^\infty \frac {g^{(k)}(\mu)}{k!} m_k 
\end{eqnarray}
where $g^{(k)}(\mu)$ denotes the $k^\text{th}$ derivative of $g$ evaluated at the expectation of $X$ and $m_k$ denotes the $k$-th central moment. For the variance and its quasi-linearity, we yield 
\begin{small}
\begin{eqnarray}\label{eq:TaylorVar}
\mathbb{V}[g(X)]  
&=& \mathbb{V} \left[    \sum_{k=0}^\infty \frac {g^{(k)}(\mu)}{k!} \, (X-\mu)^k  \right]  \nonumber \\
&=& \sum_{k=0}^\infty  \left(\frac{g^{(k)}(\mu)}{k!}\right)^2 \mathbb{V}\left[(X-\mu)^k\right]  \nonumber \\
&=& \sum_{k=0}^\infty \left(\frac{g^{(k)}(\mu)}{k!}\right)^2 (m_{2k}-m_k^2).
\end{eqnarray}
\end{small} 
\hspace{-2.5ex} By omitting terms of higher orders, we yield the linear approximations 
$\mathbb{E}[g(X)] \approx g(\mu)$ and $\mathbb{V}[g(X)] \approx g'(\mu)^2 m_1$.

So far, we have considered a smooth function with only one argument in order to guarantee an easy understanding of this methodology. When considering more arguments, we use a Taylor series in more dimensions and yield equivalent results, which, together with the assumption of normality, form the \textbf{Gaussian Error Propagation}  \cite{Ku,Bevington,Taylor}.

\subsection{Ranking Error}
Let us consider two random variables $Z_1\sim\mathcal{N}(\mu_1,\sigma_1)$ and $Z_2\sim\mathcal{N}(\mu_2,\sigma_2)$ representing arbitrary metric outcomes for two different data mining approaches. Let us additionally define the auxiliary variable
\begin{equation}
W:=(Z_1-Z_2)\sim\mathcal{N}\left(\mu_1-\mu_2,\sqrt{\sigma_1^2+\sigma_2^2}\right).
\end{equation}

The most intuitive way to build a ranking of two distributions is the comparison of their expected values.
If $\mu_1< \mu_2$, then we consider approach 1 to be better than approach 2. Due to the non-vanishing variance, this decision may be subject to an error which occurs with a probability of
\begin{equation}
P(Z_1\geq Z_2) = P(W\geq 0) = 1-F_W(0),
\end{equation}
where $F_W$ is the cumulative distribution function of $W$. Since $W$ is normally distributed, it can be represented by a transformation of the standard-normal distribution and so we can also express $F_W$ in terms of the standard-normal cumulative distribution function $\Phi$. So we finally yield
\begin{equation}
P(Z_1\geq Z_2) = \Phi\left(\frac{\mu_1-\mu_2}{\sqrt{\sigma_1^2+\sigma_2^2}}\right).
\end{equation}

\section{Application: The Probabilistic RMSE}
In this section we will derive a closed form approximation for the RMSE's probability density and therefore define 
\begin{equation} \label{eq:RMSE}
\operatorname{RMSE} = g(X_1,\ldots,X_N) := \sqrt{\tfrac{1}{N}\textstyle{\sum_{\nu}} (X_\nu - \pi_\nu)^2},
\end{equation}
where $X_\nu$ is the feedback distribution for a specific user-item-pair and $\pi_\nu$ is the corresponding prediction of an arbitrary algorithm (e.g. recommender system).

\subsection{Density Approximation}
Since we have to face multiple arguments, we would usually need a Taylor series in several variables, which is quite ugly for demonstration purposes. Therefore, we first condense all user feedbacks $X_1,\ldots,X_N$ into a single random variable and then use the one-dimensional Taylor approximation. 
In doing so, we choose Gaussians as the underlying data model. By this means, every feedback $X_\nu \sim\mathcal{N}(\mu_\nu,\sigma_\nu)$ can be written as 
$X_\nu = \sigma_\nu\mathbb{I}+\mu_\nu$ where $\mathbb{I}\sim\mathcal{N}(0,1)$. 
Hence, with $\Delta_\nu:=\mu_\nu -\pi_\nu$ being the difference between the expected rating and its corresponding prediction, the condensed quantity $Y_\nu:=(X_\nu - \pi_\nu)^2$ receives the expectation
\begin{eqnarray}
\mathbb{E}[Y_\nu] &=& \mathbb{E}[(\sigma_\nu\mathbb{I}+\Delta_\nu)^2] \nonumber 
								= \mathbb{E}[\sigma_\nu^2\mathbb{I}^2+2\mathbb{I}\sigma_\nu\pi_\nu + \Delta_\nu^2] \\
								&=& \sigma_\nu^2\mathbb{E}[\mathbb{I}^2]
								+2\sigma_\nu\pi_\nu\mathbb{E}[\mathbb{I}]+ \mathbb{E}[\Delta_\nu^2] \nonumber \\
								&=&  \sigma_\nu^2 +\Delta_\nu^2
\end{eqnarray}
as well as the variance
\begin{eqnarray}
\mathbb{V}[Y_\nu] &=& \mathbb{V}[(\sigma_\nu\mathbb{I}+\Delta_\nu)^2] \nonumber
								=  \mathbb{V}[\sigma_\nu^2\mathbb{I}^2+2\mathbb{I}\sigma_\nu\pi_\nu + \Delta_\nu^2] \\
								&=&	\sigma_\nu^4 \left(\mathbb{E}[\mathbb{I}^4]-\mathbb{E}[\mathbb{I}^2]^2 \right)
										+4\sigma_\nu^2\pi_\nu^2 \nonumber\\
								&=&			2\sigma_\nu^4+4\sigma_\nu^2\pi_\nu^2 .
\end{eqnarray}
We thus obtain a $\chi^2$-distribution for $Z:=\tfrac{1}{N}\sum_{\nu}Y_\nu$ which converges into a Gaussian for a large number $N$ of user feedback. The parameters of this Gaussian are
\begin{eqnarray}
\mathbb{E}[Z] &=& \frac{1}{N}\sum_{\nu}\mathbb{E}[Y_\nu] 
								= \frac{1}{N}\sum_{\nu} \sigma_\nu^2 +\Delta_\nu^2 \label{eq:MSEexpectation} \\
\mathbb{V}[Z] &=& \frac{1}{N^2}\sum_{\nu}\mathbb{V}[Y_\nu] 
								= \frac{2}{N^2}\sum_{\nu} \sigma_\nu^4+2\sigma_\nu^2\pi_\nu^2 . \label{eq:MSEvariance}
\end{eqnarray}
The RMSE can therefore be represented by the root function of one single condensed random variable $Z$, i.e. 
\begin{equation}
\operatorname{RMSE} = g(X_1,\ldots,X_N) \equiv h(Z):=\sqrt{Z}.
\end{equation}
Application of the one-dimensional Taylor approximations from \ref{eq:TaylorExpect} and \ref{eq:TaylorVar} lead to the parameters 
\begin{eqnarray}
\mathbb{E}[\operatorname{RMSE}] &\approx &  \sqrt{\mathbb{E}[Z]} \label{eq:RMSEexpect}
							= \sqrt{\frac{1}{N} \textstyle{\sum_{\nu}} \sigma_\nu^2+\Delta_\nu^2} \\[1ex]
\mathbb{V}[\operatorname{RMSE}] &\approx &  \frac{\mathbb{V}[Z]}{4\mathbb{E}[Z]}
							=\frac{\textstyle{\sum_{\nu}}\sigma_\nu^4+2\sigma_\nu^2\Delta_\nu^2}{2N\cdot \textstyle{\sum_{\nu}}\sigma_\nu^2+\Delta_\nu^2} \label{eq:RMSEvariance}
\end{eqnarray}
of an assumed Gaussian, which is indeed a suitable model, as we will confirm in the following section.

\subsection{Goodness of Fit}
Since we omit terms of higher orders, we only yield approximations and would therefore like to briefly discuss their quality. For this purpose, we create theoretical test data which comprises all possibilities of uncertain user feedback with respect to a given answer scale. On these records, we compare the simulated expectations and variances with the approximated ones in a regression analysis. Concerning the distribution model itself, we investigate the degree similarity using the Jensen-Shannon divergence.\\[.25ex]

\textsc{Test Data Construction.}
The degree of uncertainty, as well as the extent of deviation between a ranking and its prediction, is determined by the answer scale. Our theoretical data is constructed in accordance to the commonly used 5-star scale and we additionally assume the human uncertainty to be measured by five repeated rating trials.
Thus, the deviations are bound by $\Delta_\nu \in  [0\, , 4]$ whereas the variance can range within $\sigma_\nu^2 \in [0.16 \, , 3.86]$. For each of the following analyses, we randomly sample $N$ user-item-pairs $(\Delta_\nu,\sigma_\nu^2)$ uniformly from their intervals and perform further computations.\\[.25ex]

\begin{figure}[t]
    \centering
    \begin{subfigure}{0.235\textwidth}
        \includegraphics[width=\textwidth]{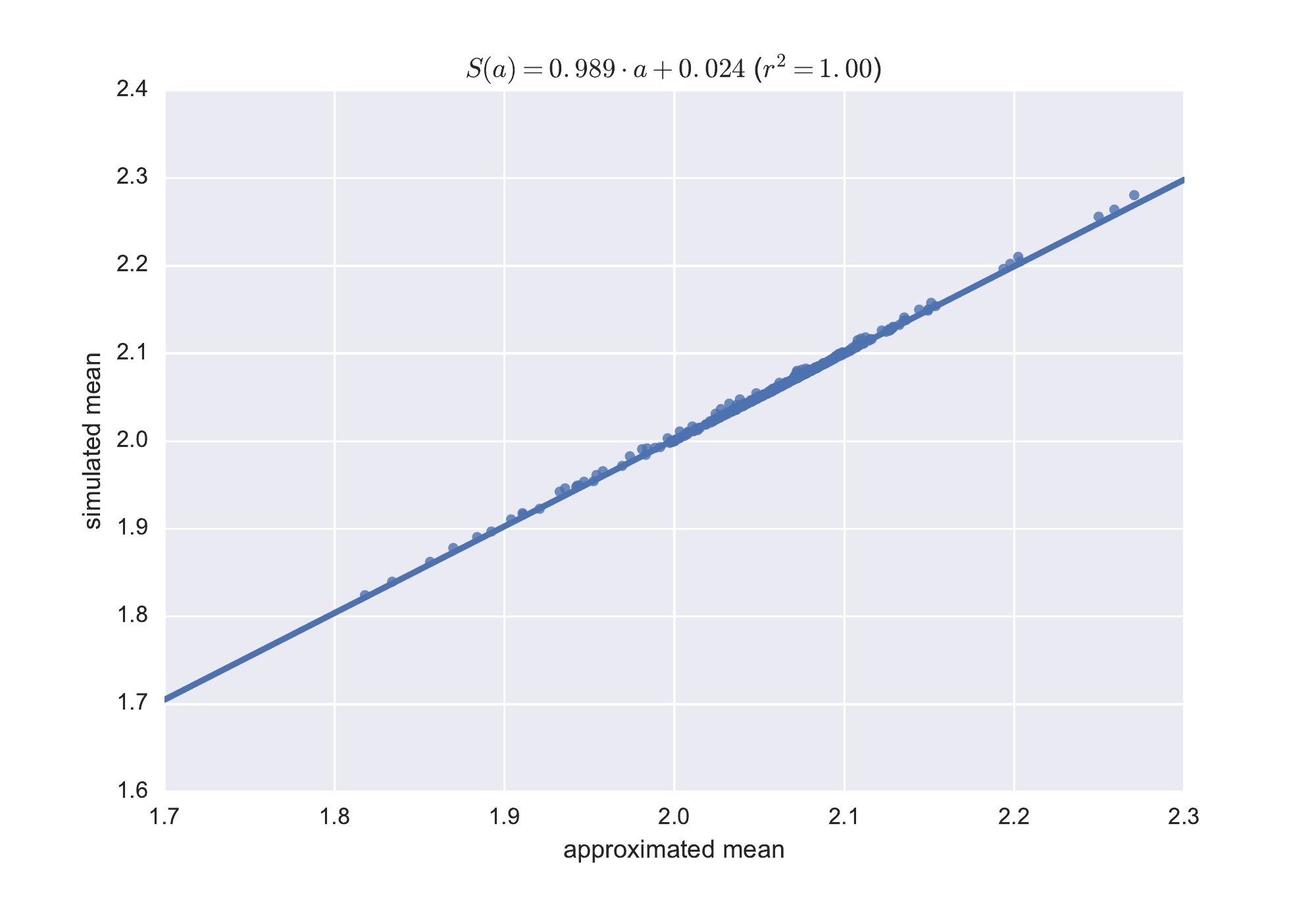}
        \caption{Regression for approximated against simulated means}
    \end{subfigure}
    \begin{subfigure}{0.235\textwidth}
        \includegraphics[width=\textwidth]{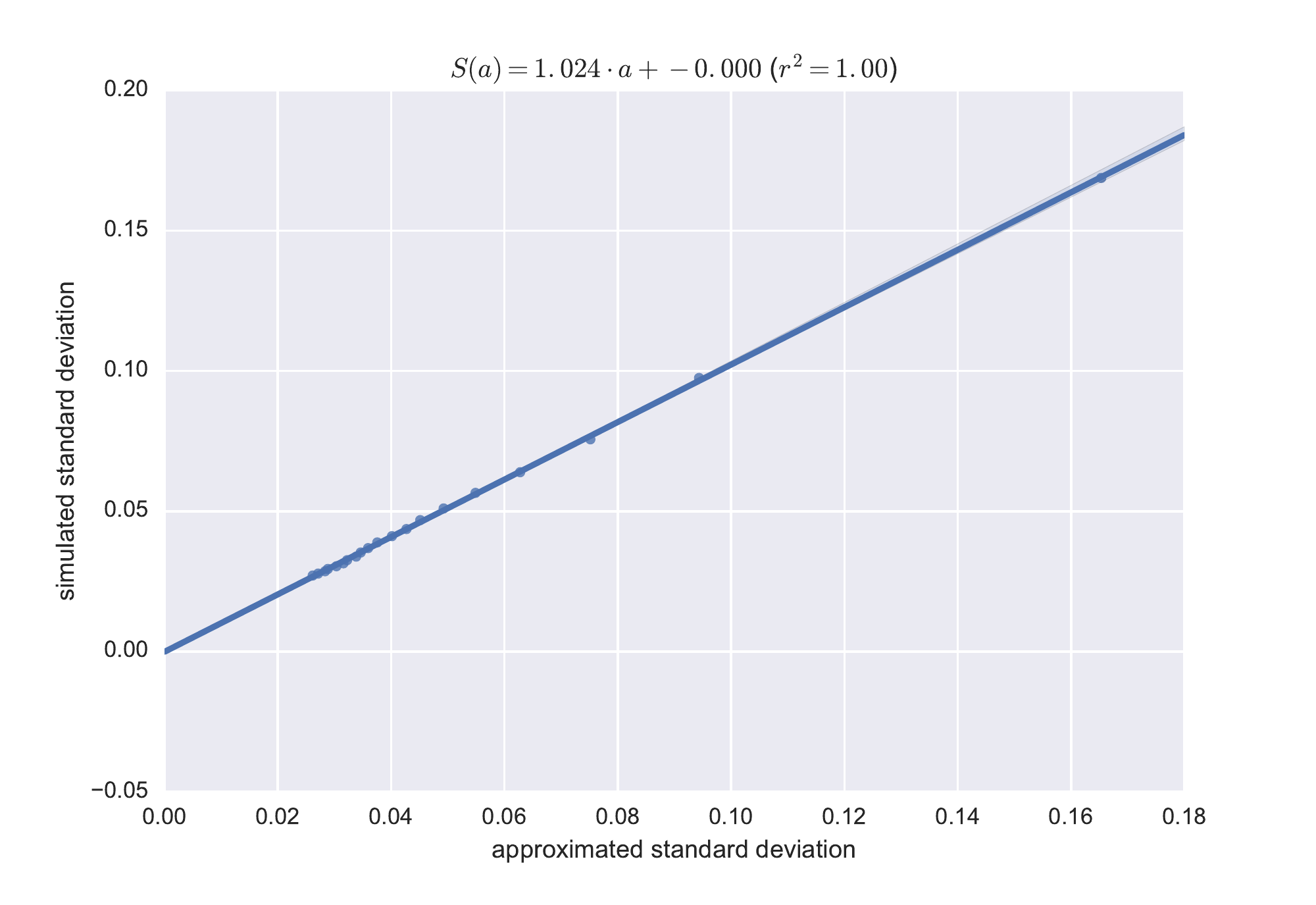}
        \caption{Regression for approximated against simulated variances}
    \end{subfigure}
    \begin{subfigure}{0.5\textwidth}
        \includegraphics[width=\textwidth]{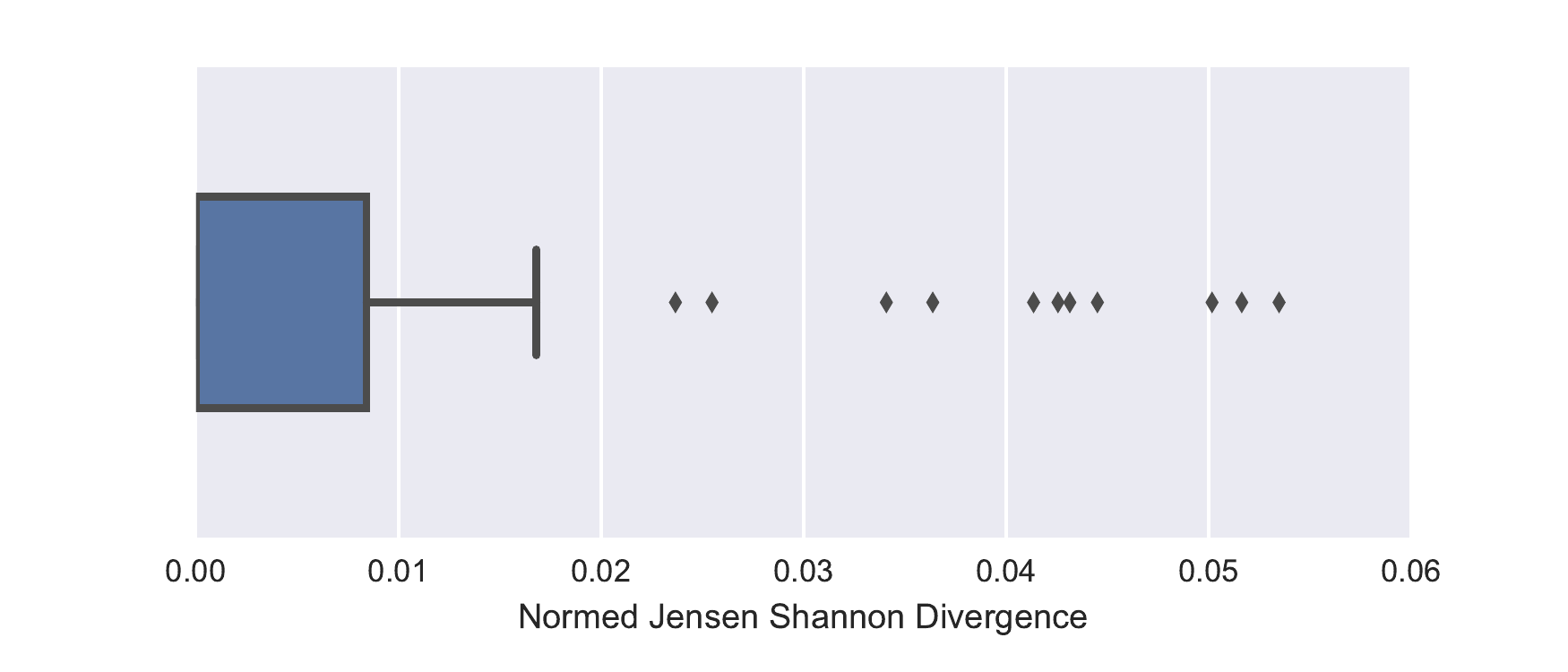}
        \caption{nJSD of approximated against simulated distributions}
    \end{subfigure}
    \caption{Goodness of fit for the RMSE approximations}
    \label{fig:GoF}
\end{figure}

\textsc{Parameter Matching.}
To verify the mean and variance of the RMSE, we gradually fix a particular number 
$N\in \{50, 250, 500, \ldots, 2500\}$ of user-item-pairs and construct the test data as described above. On this basis it is just straightforward to compute these parameters by \ref{eq:RMSEexpect} and \ref{eq:RMSEvariance} (subscript \textit{apr}) or respectively deduce them from a Monte-Carlo-Sample (subscript \textit{sim}). When repeating this step 50 times for each $N$, one obtains $2\,500$ data points $(\mu_{apr},\mu_{sim})$ and $(\sigma_{apr}^2,\sigma_{sim}^2)$ to be analysed by linear regression. In case of perfect matching, plotting approximated parameters against simulated ones should result into a straight line with slope equal to 1, intercept equal to 0 and correlation coefficient equal to 1. The results
\begin{eqnarray}
\mu_{sim} &=& 0.99 \cdot \mu_{apr} + 0.02  \qquad (r^2=1.00)\\
\sigma_{sim}^2 &=& 1.02 \cdot \sigma_{apr}^2 + 0.00 \qquad (r^2=1.00)
\end{eqnarray}
(and also depicted in Figure \ref{fig:GoF}) show that this conditions are almost fully achieved. 
Hence, our approximations are surprisingly good.

\begin{figure*}
    \centering
    \begin{subfigure}{0.329\textwidth}
        \includegraphics[width=\textwidth]{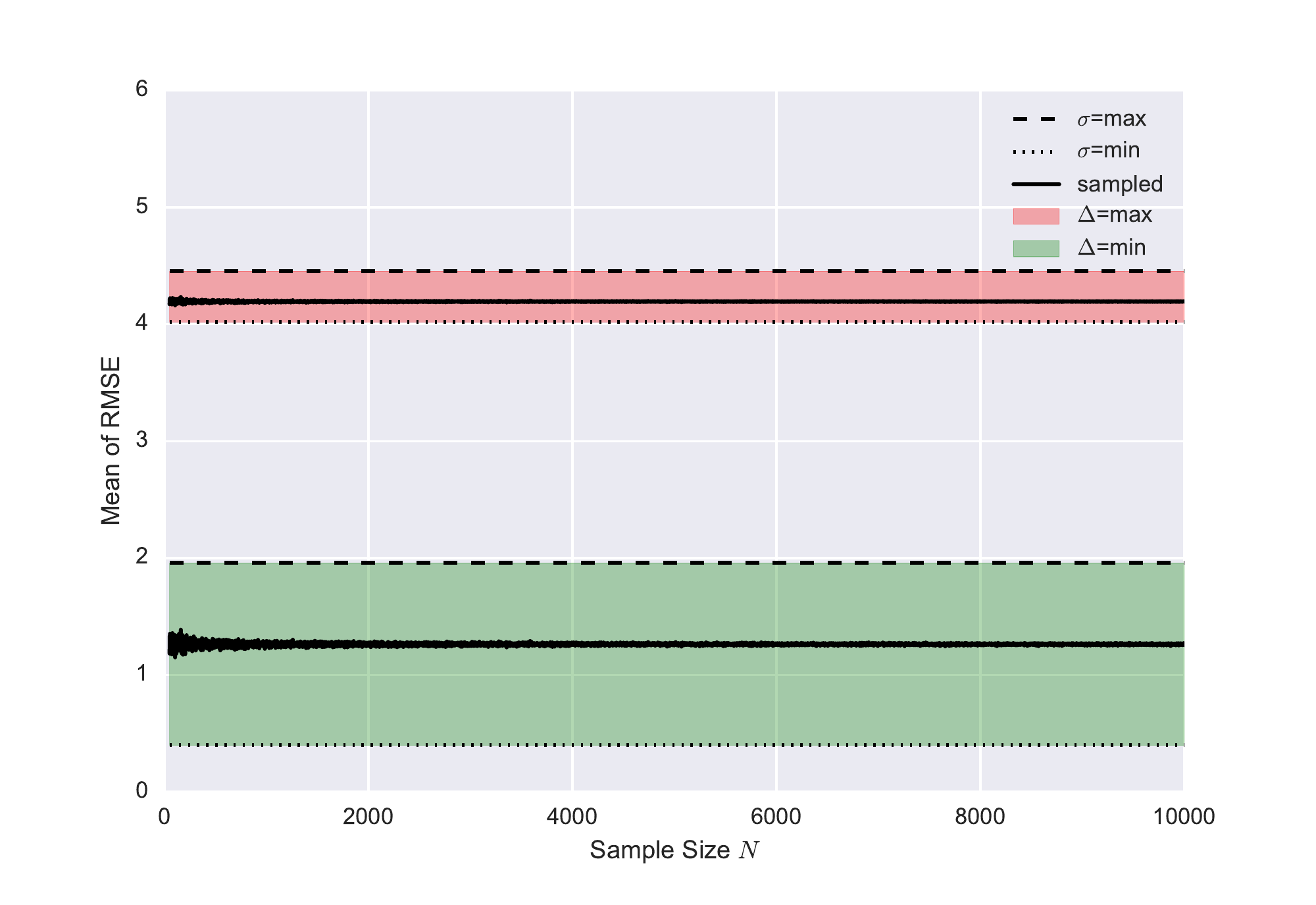}
        \caption{$\mu_{\text{RMSE}}$ with respect to $N$}
    \end{subfigure}
    \hfill
    \begin{subfigure}{0.329\textwidth}
        \includegraphics[width=\textwidth]{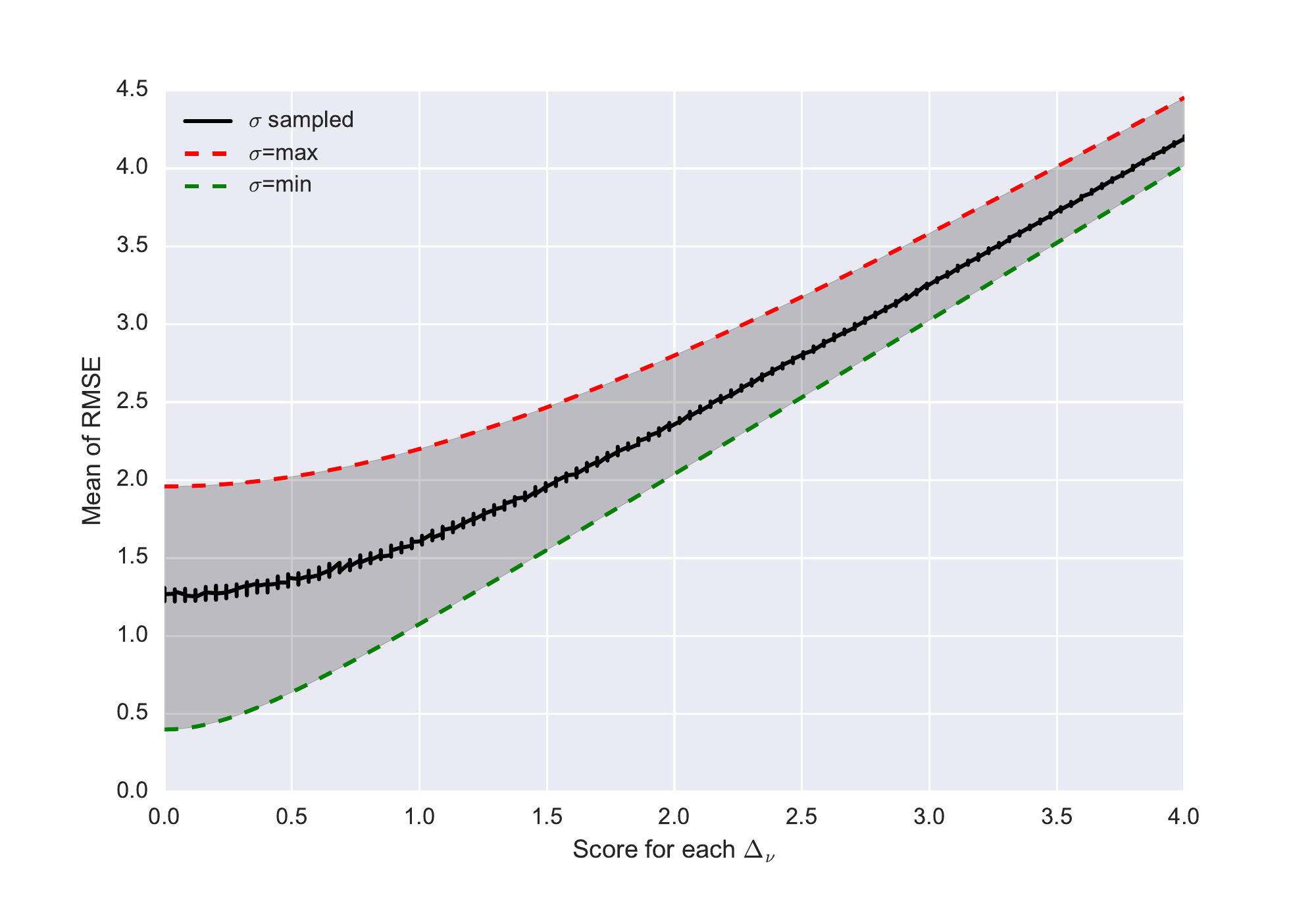}
        \caption{$\mu_{\text{RMSE}}$  with respect to $\Delta_\nu$}
    \end{subfigure}
    \hfill
    \begin{subfigure}{0.329\textwidth}
        \includegraphics[width=\textwidth]{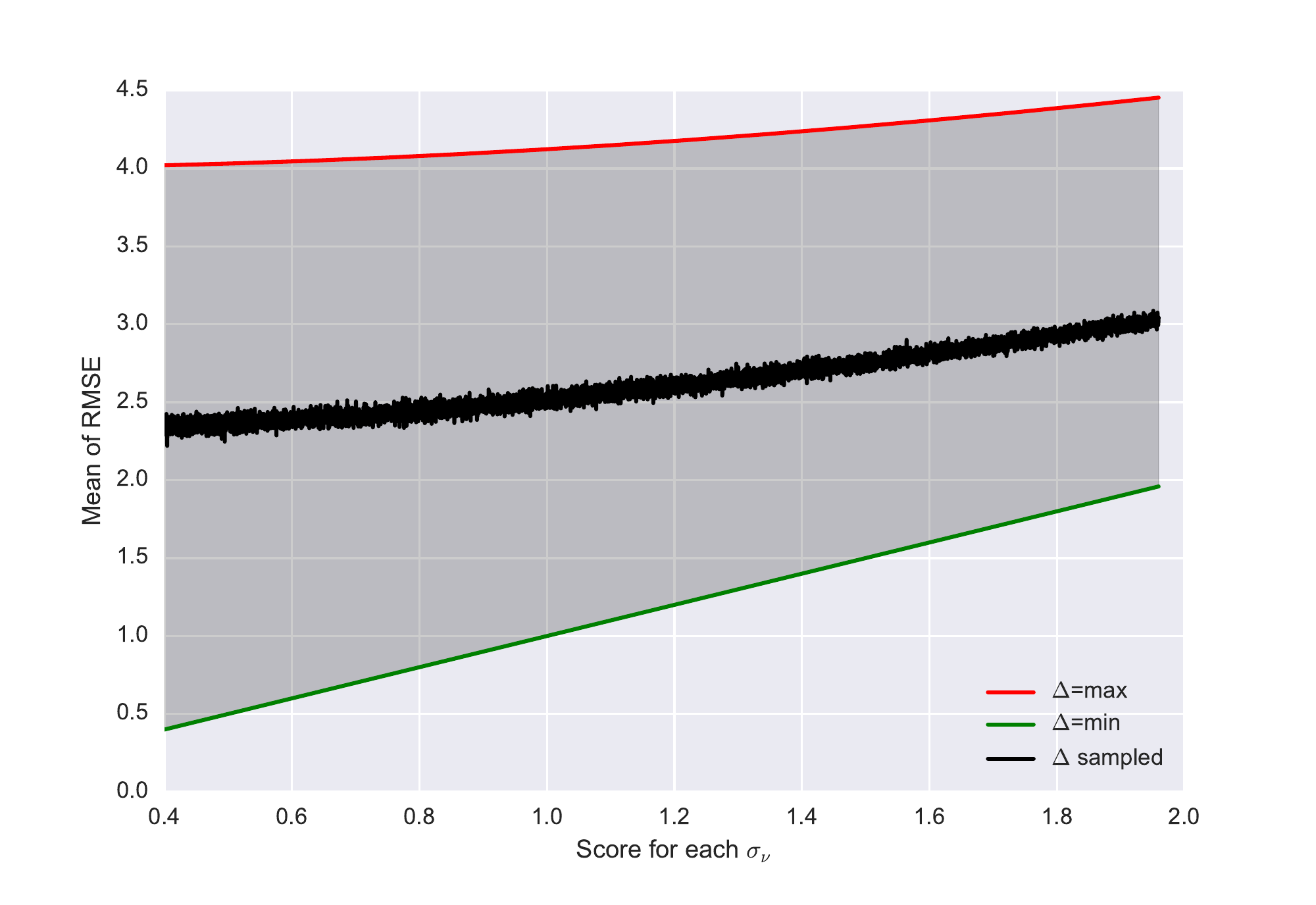}
        \caption{$\mu_{\text{RMSE}}$ with respect to $\sigma_\nu^2$}
    \end{subfigure}
    \caption{Sensitivity analysis for the expectation $\mu_{\text{RMSE}}$ of the RMSE distribution.}
    \label{fig:SensMean}
\end{figure*}

\textsc{Distribution Similarity.}
Even if both parameters appear to have a good matching, this doesn't prove the assumption of a Gaussian. For this, we consider the Jensen-Shannon divergence (JSD) between the probability distribution $P_ {sim}$ from Monte-Carlo-Simulation and our mathematically approximated distribution $P_ {apr}$ via
\begin{eqnarray}
\operatorname{JSD}(P_{sim}| P_{apr}) \nonumber
&=& \frac{D_\mathrm{KL}(P_{sim}| M)}{2}  + \frac{D_\mathrm{KL}(P_{apr}| M)}{2} \\
\text{with } M &:=& (P_{sim}+P_{apr})/2
\end{eqnarray}
where $D_\mathrm{KL}(P_1|P_2)$ is the Kullback-Leibler divergence. The JSD has the boundaries
\begin{equation}
0 \leq \operatorname{JSD} \leq 2\log(2)  \quad\text{or}\quad 0 \leq \tfrac{\operatorname{JSD}}{2\log(2)} \leq 1 .
\end{equation}
We denote the quantity in the right inequality as the normed Jensen-Shannon divergence (nJSD). It is a common measure for the similarity of two probability distributions, where zero represents a perfect matching. Figure \ref{fig:GoF}c depicts the different outcomes by using of a box-plot. The lower whisker, as well as the first quartile and the median are altogether approximately zero. Additionally, the third quartile (75\% of nJSD values) is below $0.02$ and even the outliers have very low values that do not exceed $0.06$. This strongly supports the assumption of a Gaussian.

\subsection{Proof of Uncertainty Impact}
Now having confirmed our model to be an adequate representation of human uncertainty and its propagation, we may use this instrument to reveal relevant knowledge and insights that are related to the field of data mining.

Concerning the expected value of the RMSE in \ref{eq:RMSEexpect}, we notice a heavy impact of the human uncertainty, i.e. it influences the expected value with same potency as the deviations themselves.
This is a quite astonishing finding, since one would have expected the additional uncertainty of user feedback to just induce some uncertainty of the utilised metric. Put simply: One might want to maintain the metric score (current status quo) and simply add some additional variance to it. Equation \ref{eq:RMSEexpect} however reveals that this easy way of thinking is not adequate, because human uncertainty leads to a remarkable shift of the expectation. Unfortunately, it is not easy to subtract out this bias, as the sum can not be decomposed into separate terms due to non-linearity of the root function.

Concerning the variance of the RMSE in \ref{eq:RMSEvariance}, we recognise each uncertainty to contribute significantly (i.e. to the power of 4). Also the deviances between feedback and prediction contribute to the variance and thus influences the precision of the RMSE. 
Overall, we see that human uncertainty is either equally potent or higher potent than the actual deviances. 
But since each parameter of the resulting RMSE distribution depends on both influential factors at the same time (deviances and uncertainty), it is difficult to discover the sole impact of human uncertainty.
Additionally, the possible extent of human uncertainty can be of a different magnitude than the deviations themselves. Therefore, although a significant impact of human uncertainty can be considered to be proven, a more differentiated elaboration of its impact is necessary at this point.

\begin{figure*}
    \centering
    \begin{subfigure}{0.329\textwidth}
        \includegraphics[width=\textwidth]{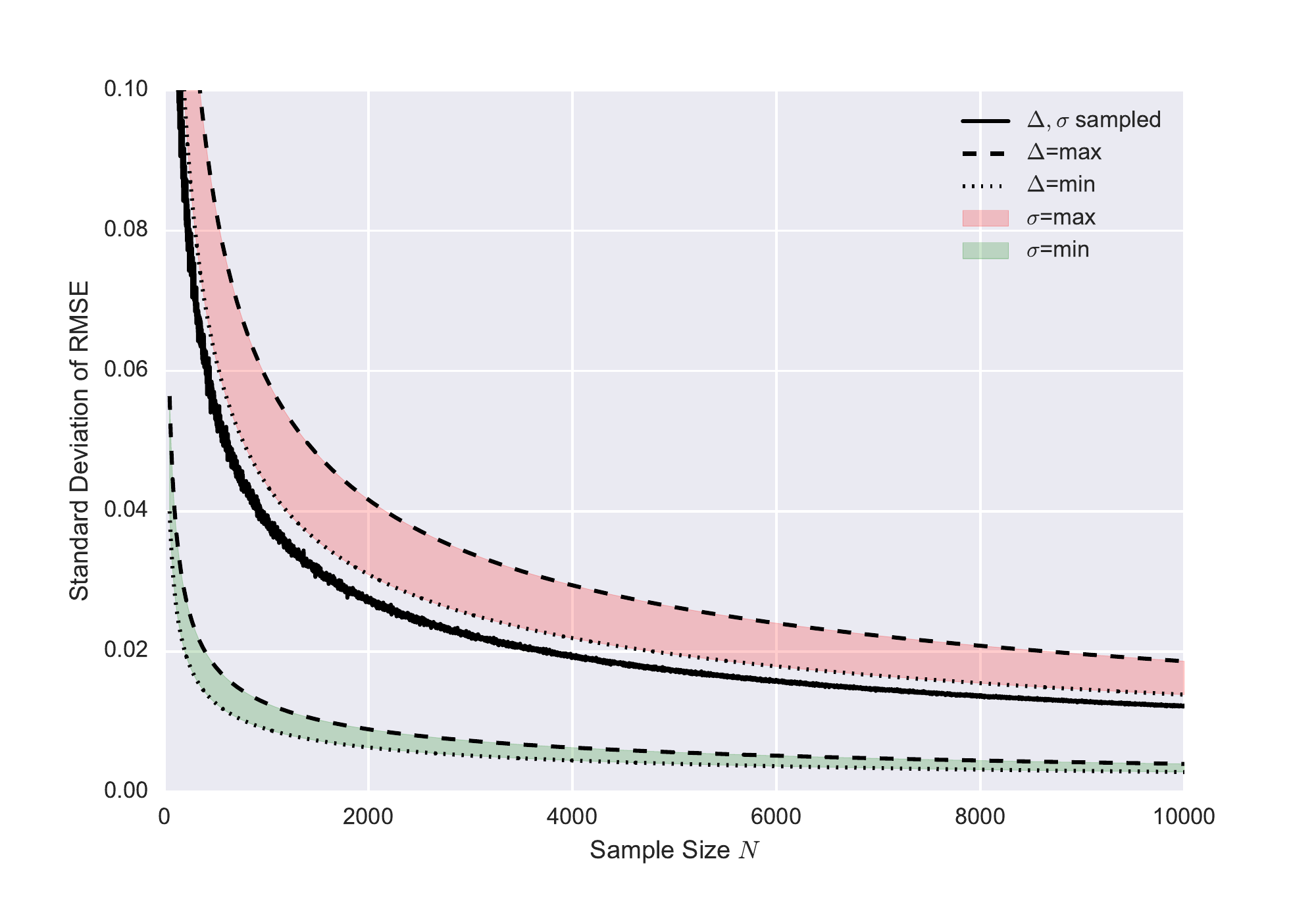}
        \caption{$\sigma_{\text{RMSE}}$ with respect to $N$}
    \end{subfigure}
    \hfill
    \begin{subfigure}{0.329\textwidth}
        \includegraphics[width=\textwidth]{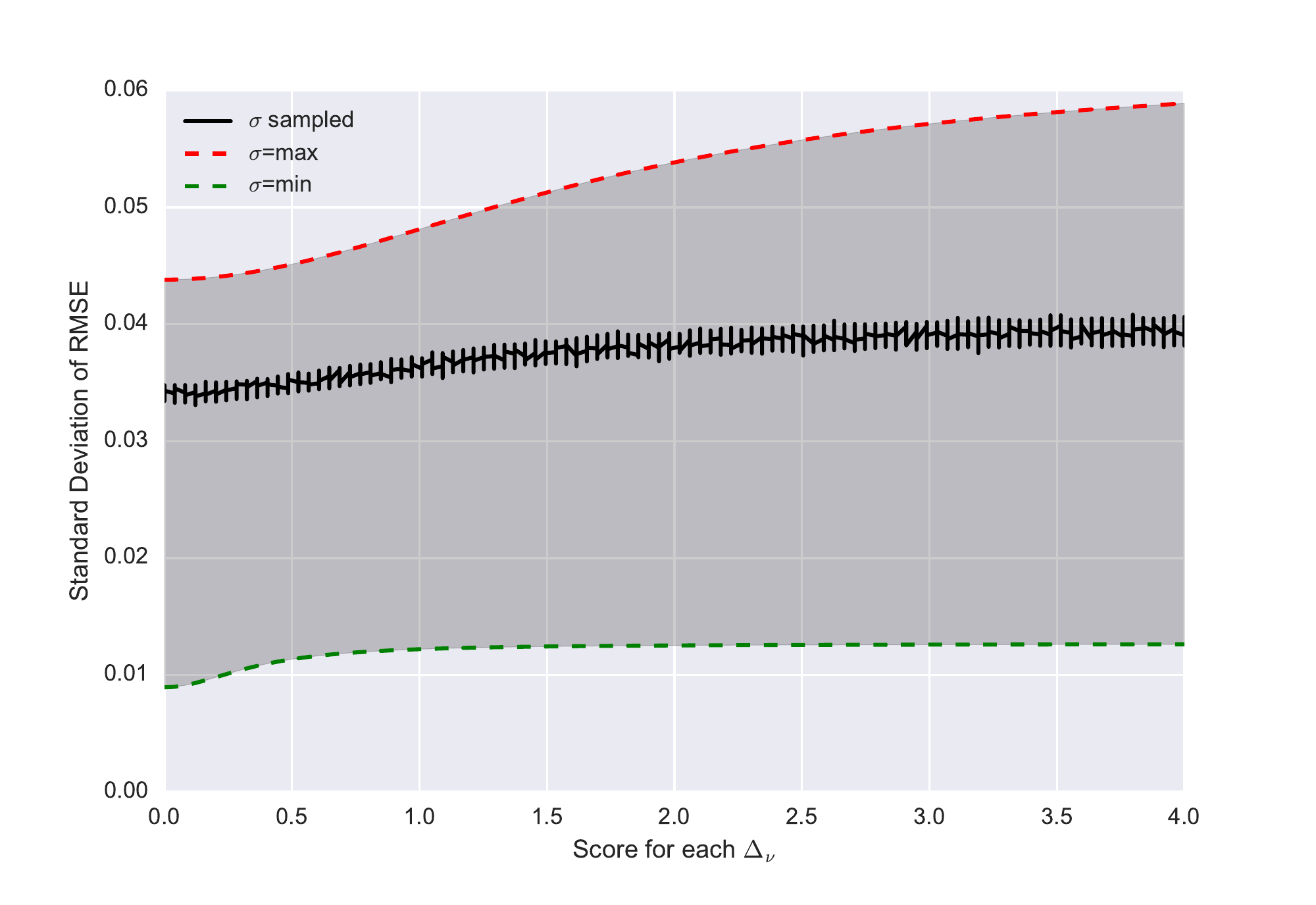}
        \caption{$\sigma_{\text{RMSE}}$  with respect to $\Delta_\nu$}
    \end{subfigure}
    \hfill
    \begin{subfigure}{0.329\textwidth}
        \includegraphics[width=\textwidth]{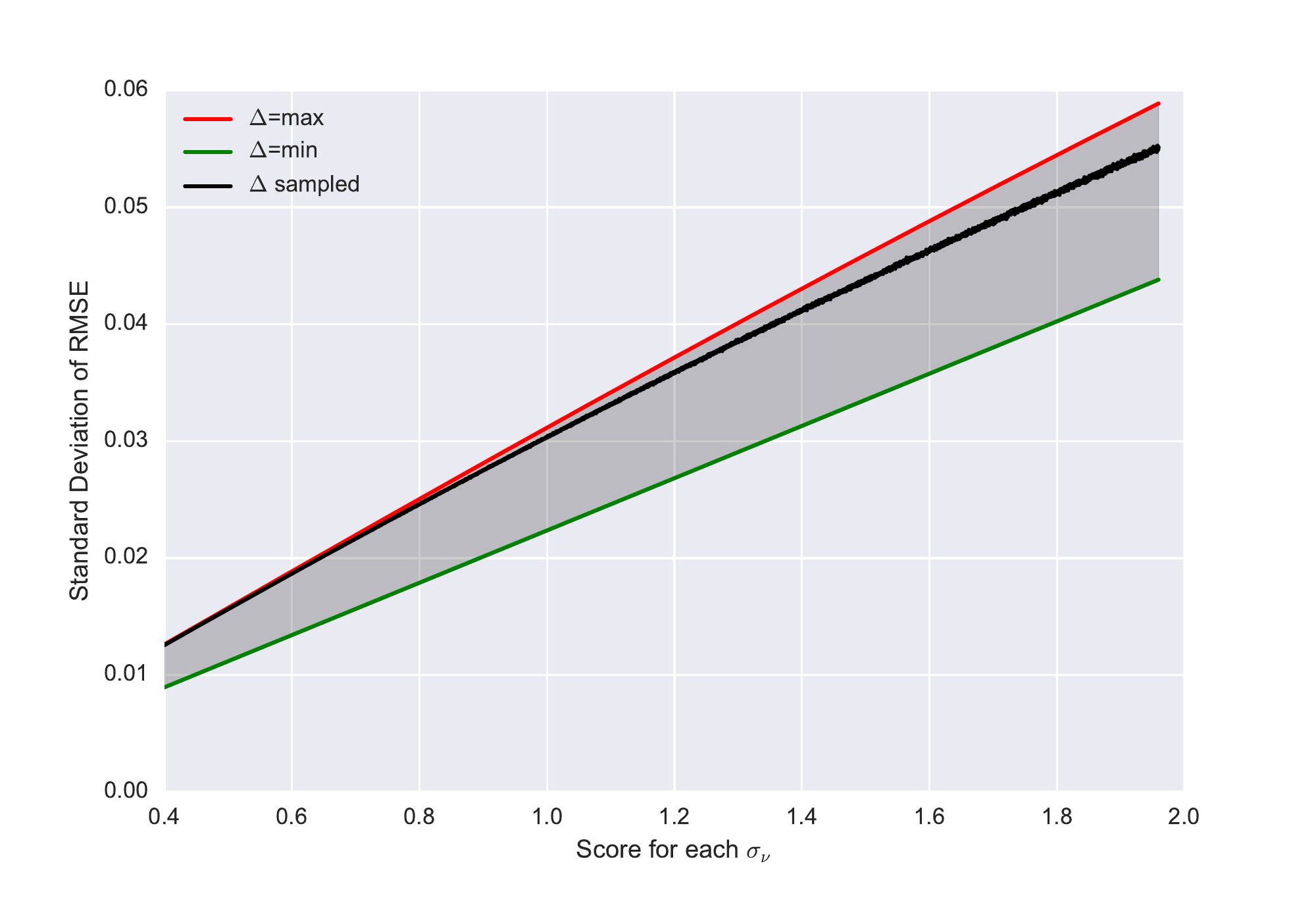}
        \caption{$\sigma_{\text{RMSE}}$ with respect to $\sigma_\nu^2$}
    \end{subfigure}
    \caption{Sensitivity analysis for the variance $\sigma_{\text{RMSE}}^2$ of the RMSE distribution.}
    \label{fig:SensSTD}
\end{figure*}

\subsection{Understanding the Impact}

In this section, we will take a closer look at the properties of the RMSE distribution. For this purpose, the individual dependencies and their effects are analysed in a sensitivity analysis, i.e. we determine how a particular distribution parameter responds to the variation of its arguments. Therefore, we vary one argument within its reasonable boundaries while fixing all the other arguments at the same time.\\[.25ex]

\textsc{Impact on the Metric's Mean.}
Figure \ref{fig:SensMean} depicts the outcomes for the expectation 
$\mu_{\text{RMSE}}$ of the RMSE in correspondence with the number $N$ of user-item-pairs, the average deviance $\Delta_\nu$ between feedback and prediction as well as the the average human uncertainty $\sigma_\nu^2$.

Subfigure \ref{fig:SensMean}a shows that the mean of the RMSE is not influenced by $N$ (straight line with vanishing slope). We also recognise that the impact of human uncertainty is much greater for small deviations than for large ones (width of coloured bands). For small deviations (green/lower band) in particular, human uncertainty may shift the RMSE's location from $0.5$ to $2.0$ (+300\%), whereas a shift can only increase values from $4.0$ up to $4.5$ (+13\%) for large deviations (red/upper band).
This can be explained by the fact that although both quantities are equal in magnitude, the differences are much greater for the deviations than for the human uncertainty. Thus, for large deviations, the additional contribution of human uncertainty is lower.
However, this confronts us with serious problems, because the better our data mining approaches become (lower deviances), the more impact is given to human uncertainty - and this uncertainty cannot be improved, since its origin is the cognitive process.

Subfigure \ref{fig:SensMean}b shows the reaction of the RMSE's mean on variation of the average deviation between prediction and user feedback.The curve clearly shows that there is a functional dependency with asymptote $a(x)=x$. Here, the width of the gray band is an indicator for the influence of human uncertainty, which fades for large deviations. Again, we can identify the possible shift of the expected value, but for a much finer gradation than in the figure before.

Subfigure \ref{fig:SensMean}c depicts the dependency on the average human uncertainty. At first glance, the curve looks different from the curve we had obtained for the average deviation, although both quantities contribute equally to the mean of the RMSE. This vividly demonstrates that the magnitude of human uncertainty is much more limited than that of the deviations. Indeed, this curve looks like a large zoom on the beginning part of the graph in Figure \ref{fig:SensMean}b.

Let us briefly restate: Generally, the mean of our quality metric is mainly determined by the deviations themselves. So far, this is a very good sign for our community, because what we actually wanted to measure is indeed measured - well, almost! However, for good systems, i.e. those with small deviations between predicted action and actually executed action, the human uncertainty receives an influence of up to 300\% (on a 5-star scale and along with the RMSE). Technically spoken: the better our systems become, the more impact is given to random fluctuations that occur for assessment repetitions. This is sadly not optimal for our community and is little considered in latest assessments.
It is also striking that the RMSE, as minimal as it gets, does never vanish. That is, the mere existence of human uncertainty generates an offset, i.e. an RMSE score that cannot be fallen below.
The existence of such a barrier has already been predicted in \cite{Herlocker} and is denoted as Magic Barrier. For the RMSE in particular, this Magic Barrier has been theoretically calculated in \cite{MagicBarrier}.\\[.25ex]

\textsc{Impact on the Metric's Variance.}
Figure \ref{fig:SensSTD} depicts the outcomes for the variance 
$\sigma_{\text{RMSE}}^2$ of the RMSE in correspondence with $N$, $\Delta_\nu$ and$\sigma_\nu^2$.

In Subfigure \ref{fig:SensSTD}a, we recognise the great influence of human uncertainty (a large range between the coloured bands).
Although the deviations between prediction and action also have an impact on the metric's variance, it is relatively weak and is itself dependent on the human uncertainty. That is, the impact of deviations is poor (width of the green/lower band) for a small uncertainty, but can be amplified by large uncertainties (width of the red/upper band). 
The most striking dependency of the metric's variance (precision) is the dependence on the number $N$ of user-item pairs. On the one hand it is surprising that the precision of a particular metric gains from adding more data with additional uncertainty. On the other hand we know from \ref{eq:RMSEvariance} that the variance scales with $1/N^2$. This fortunately means that we yield a gain in precision for larger data sets very quickly. This unfortunately also means that the increase in precision for even larger data sets rapidly fades. The decrease in the variance (gain in precision) up to $N=3\,000$ is tremendous (-133\%). Thereafter, a further enlargement of the data record no longer leads to a such a remarkable precision gain anymore. This finding may hold consequences on the economics of smaller studies (e.g. testing of new interfaces), since there is a point from which on additional participants will cost money but will not bring much benefit. This fast convergence means that we still have to deal with influential variances for big data.

Subfigure \ref{fig:SensSTD}b depicts the influence of the deviances.
It is apparent that there is only a weak dependency (borders are approximately straight lines with vanishing slope). For example, the magnitude of these deviations can increase the variance of the RMSE from $0.045$ to $0.06$ (+33\%, difference of red/upper curve representing high human uncertainty) or from $0.01$ to $0.012$ (+20\%, difference of green/lower curve representing low human uncertainty). In contrast, the human uncertainty itself may impact the variance much stronger (+300\%, width of gray band).
Subfigure \ref{fig:SensSTD}c demonstrates the impact of human uncertainty. We recognise linear growth of the variance with respect to human uncertainty, which is amplified for large prediction-rating-deviations (slope of the red/upper line compared to the green/lower line). At the same time, human uncertainty can increase the variance of the RMSE tremendously (difference in height of the coloured/outer lines) in comparison to the prediction-rating-deviations (width of gray band).

In summary, the variance of the RMSE is affected by deviations (small impact), by human uncertainty (large impact) as well as by the number of user-item-pairs (massive impact). This might be once again a very good sign for our community: Since we cannot improve human uncertainty, one possible way of dealing with its impact is simply to use big data. However, this way of thinking works only within certain limits, as the precision gain itself quickly decreases with increasing amount of additional data. Here we have to find the golden mean between the necessary precision and monetary expense induced by more data. It has recently been shown that even for large data records (e.g. the Netflix Prize), there is yet still a considerable variance that corresponds to different RMSE outcomes \cite{OwnMagicBarrier}. The disagreeable consequence about this non-vanishing variance is, that it is inducing a probability of error whenever we built a ranking of systems with respect to a particular metric. This has to be investigated more closely.\\[.25ex]

\textsc{Impact on the Probability of Error.}
In order to repeat the idea of error probability, we must remember that every time we build a ranking based on the location of two distributions with respect to a given metric, the non-vanishing variance may lead to an overlap of these distributions. If this overlap is too large, it may occur that specific draws from this distributions may invert the ranking order. When this error probability is too high, there is no sufficient evidence whether to assume $\operatorname{Sys}_1<\operatorname{Sys}_2$ or $\operatorname{Sys}_2<\operatorname{Sys}_1$ and both systems become undistinguishable by means of a ranking. Such a probability of error naturally respects the mean of both metric distributions (impacted by the deviations and human uncertainty) as well as their variance (impacted by number of data). For investigation of these quantities' impact, we specified two recommender systems by defining their prediction-rating-deviances, in such a way that one system's mean is constantly 10\% better than the other system's mean. We had chosen this constant difference in accordance to the Netflix Prize where improvements had to be at least 10\%. Other choices for this constant difference will produce same general results as presented below. 

Subfigure \ref{fig:SensErr}a depicts the error probability for two RMSE distributions with respect to the average deviation of prediction and rating together with the impact of the amount $N$ of data. Here we choose $P=0.05$ for being the borderline of distinguishability (in accordance to the significance levels for hypothesis testing). It is astonishing that the error curves are no constant lines, meaning that the distibguishability is different for two good recommender than for two bad ones, even though the difference of the RMSE's mean remains exactly the same.
Moreover, we see that for $ N = 50 $ no system can be distinguished from another without making a mistake in less than 5\% of all cases (blue/upper line). For $ N = 100 $ for example, only poor systems ($\Delta>3.2$) can be distinguished with an error probability of less than 5\%. Good systems ($\Delta<1$), on the other hand, can only be sufficiently distinguished with at least $2\,500$ user-item-pairs. It is shown qualitatively, that the influence of the data size has got as much influence on the distinguishability as on the variance itself.
Even the convergence of the variance for $n\to\infty$ can be found, represented by the fading distance of the curves relative to one another for increasing size of data. This supports the assumption that there is a final error curve, which is approached for large amounts of data.

Subfigure \ref{fig:SensErr}b depicts the error probability for two RMSE distributions with respect to the average deviation together with the impact of human uncertainty for a fixed data size of $N=1\,000$. Yet again we see no constant lines. It turns out that human uncertainty  can significantly shift the border of sufficient distingusihability.
For a low uncertainty (green/lower curve) extremely good systems ($\Delta<0.5$) can be brought into a ranking with only low probability of error. For a high uncertainty (red/upper curve), only medium-quality systems ($\Delta\approx 2$) can be distinguished by means of low-error rankings.

In conclusion, the impact of human uncertainty on the distinguishability is remarkable, but gives the impression of not being as striking as the the impact of the data size. At this point we have to emphasise once again that it is the mere existence of human uncertainty that causes all this problems about ranking error and distinguishability. 
Even with big data we cannot completely get rid of this problem, as the gain of distinguishability fades for additional data. Another surprising fact is, that two systems with a relative difference of 10\% by means of the RMSE can be distinguished and put into a ranking order, if these are low-quality systems. On the other hand, these can not be distinguished anymore if they are high-quality, although the relative difference is still 10\%. This indicates that the better our systems become, the more additional improvement a superior system needs in order for it to be recognised as such with statistical evidence.

\begin{figure}
    \centering
    \begin{subfigure}{0.49\textwidth}
        \includegraphics[width=\textwidth]{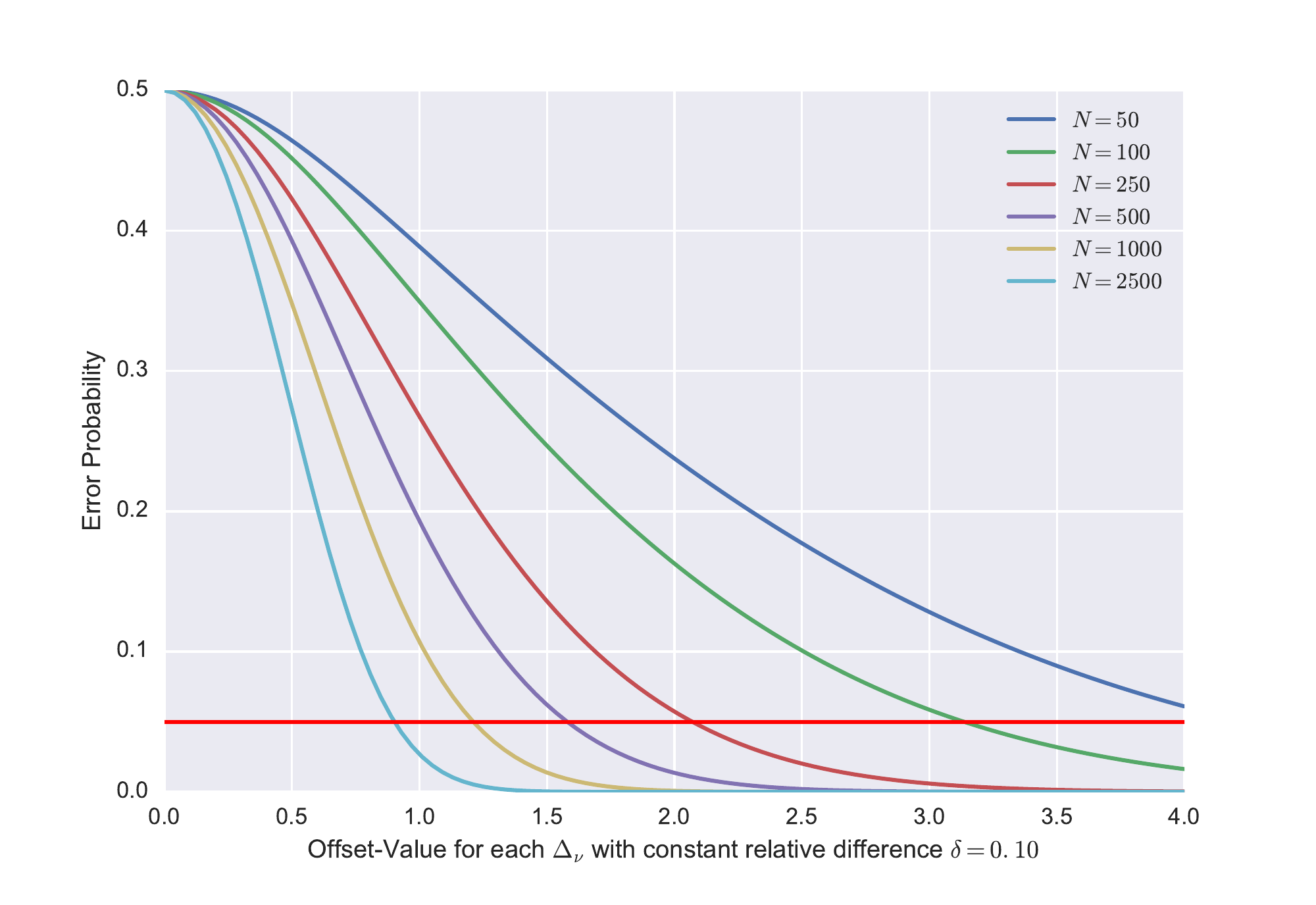}
        \caption{Error Probability with respect to $N$}
    \end{subfigure}
    \begin{subfigure}{0.49\textwidth}
        \includegraphics[width=\textwidth]{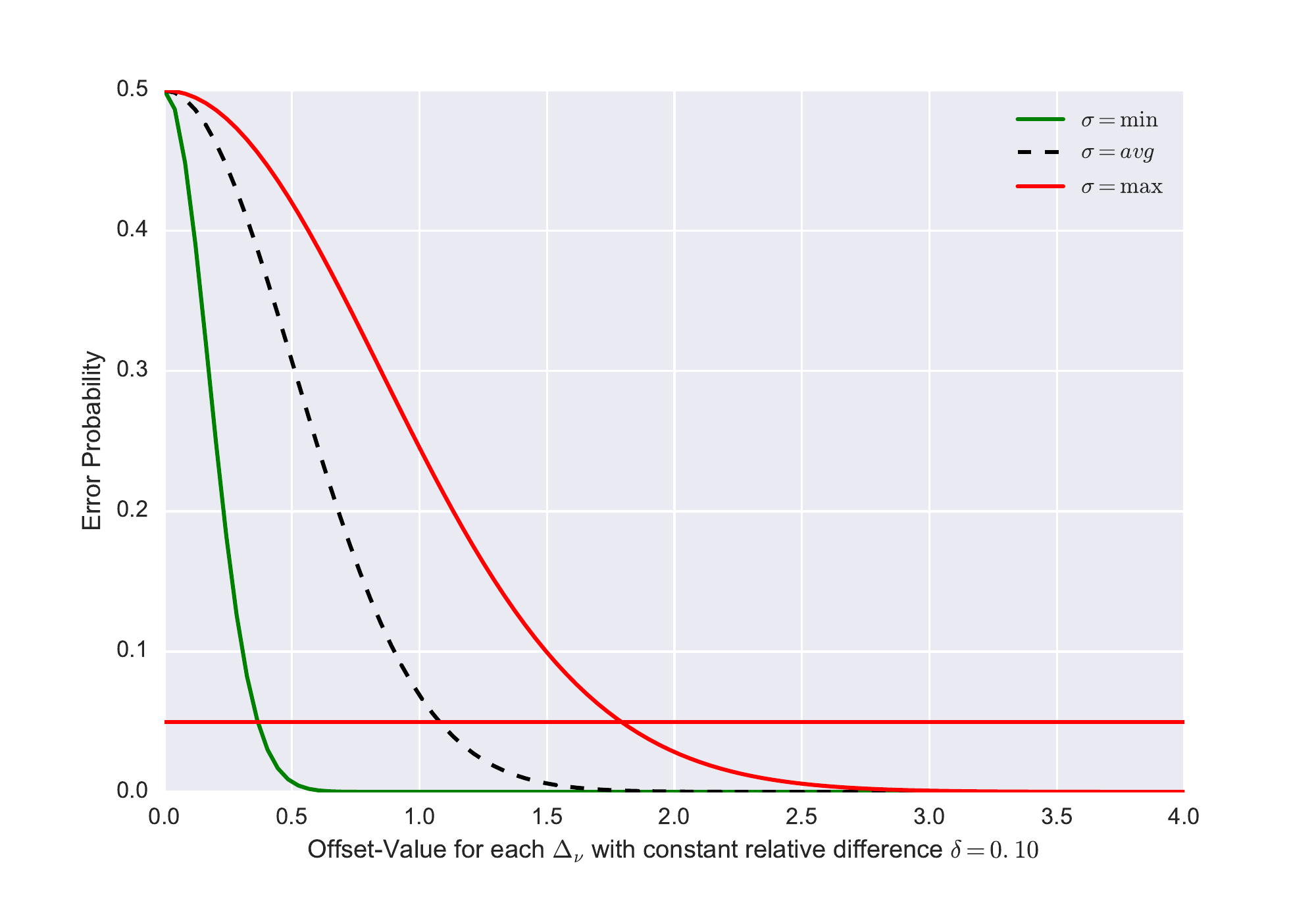}
        \caption{Error Probability with respect to $\sigma_\nu$}
    \end{subfigure}
    \caption{Sensitivity analysis for the error probability.}
    \label{fig:SensErr}
\end{figure}

\section{Solution Strategies}
The consequences of our findings so far are not very pleasing. One possible solution, supported by Subfigures \ref{fig:SensSTD}a and \ref{fig:SensErr}a, is simply to use big data. It is common knowledge that big data leads to a better accuracy, but however, our results show that even big data cannot represent a complete solution since the precision gain for a given quality metric is fading very quickly. 
In additional research, we were able to show that even for the Netflix Prize test record ($N = 2.8\cdot 10^6$) the variance of the RMSE is still affecting the Top 10 ranking, i.e. some placements are subject to very high probabilities of error\cite{OwnMagicBarrier}. So, other solutions are needed.

Our idea is simply to modify existing quality measures slightly to make them sensitive to fluctuations of user feedback. In other words, each time a rating is compared to a model-based prediction, we must examine whether the observed deviations are significant or just in nature of human uncertainty. In doing so, we divide the set of all deviations into two subsets: One subset contains all the deviations around a predictor $\pi_\nu$ that can be considered as induced by human uncertainty. The other subset contains all deviations whose extent cannot be explained by this uncertainty and thus seems to be induced by the prediction model itself. In this case, it deems viable to calculate the quality metric by taking into account only those deviations that are related to the algorithm rather than to human uncertainty. 

\begin{figure}[t]
    \centering
        \includegraphics[width=\linewidth]{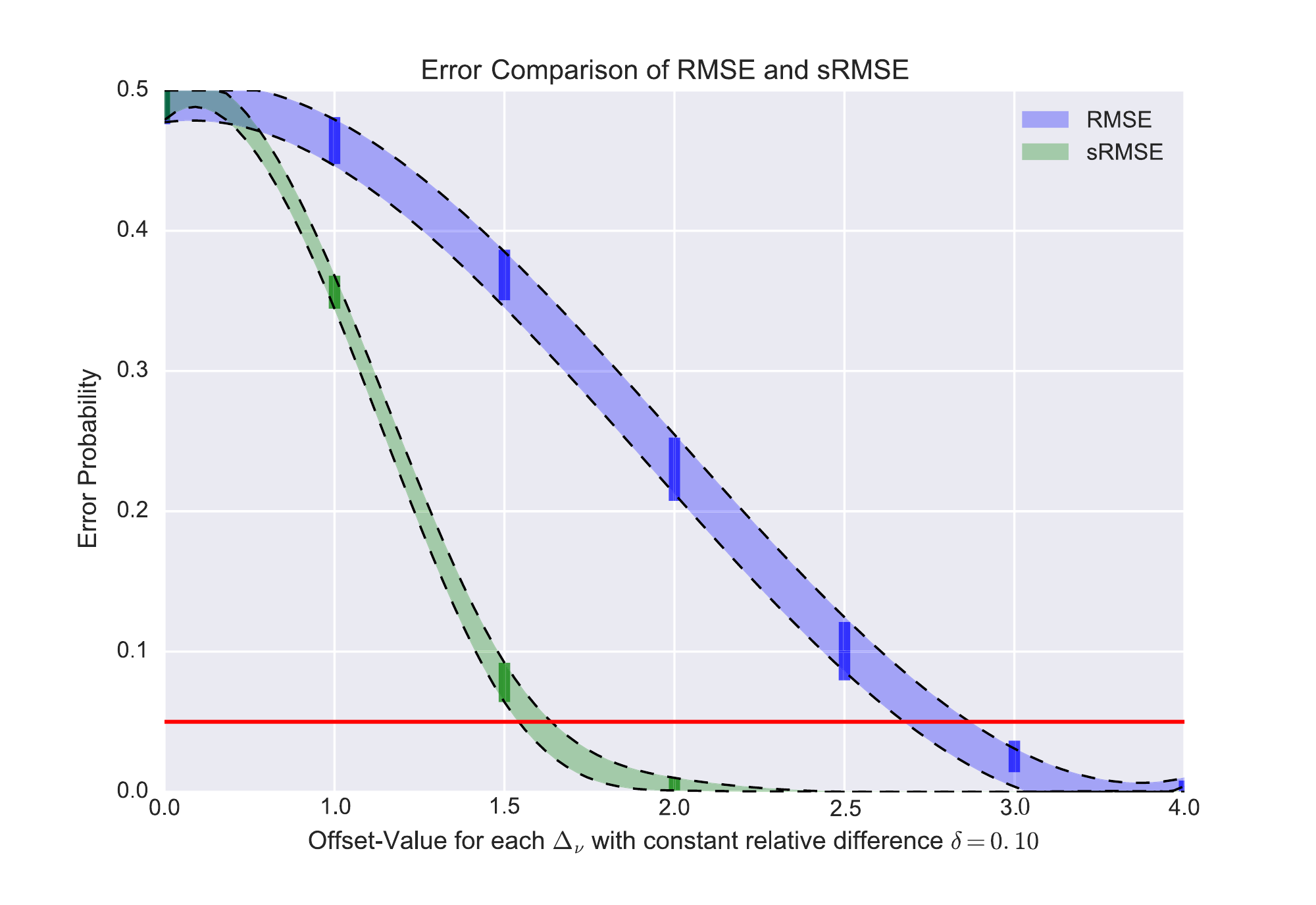}
    \caption{Comparison of the error probabilities when calculated on the RMSE and sRMSE respectively.}
    \label{fig:ErrSRMSE}
\end{figure}

We will exemplify this idea with the RMSE. Following the explanation above, we have to use statistical hypothesis testing to decide whether a realisation $r_\nu$ of a feedback distribution $R_\nu$ is equal to a model-based prediction $\pi_\nu$ or not. In mathematical notation, we have to test
\begin{equation}
H_0 \colon  r_\nu = \pi_\nu \quad\textbf{vs.}\quad H_1 \colon  r_\nu \neq \pi_\nu
\end{equation}
at a given significance level $\alpha$.
For known density functions, the region of rejection can be constructed as the complement of $I_{\alpha} = [ \pi_\nu - a  \,;\, \pi_\nu + a]$ where $a$ is chosen such that
\begin{equation}\label{eq:confinterval}
\int_{\pi_\nu - a}^{\pi_\nu + a} f_{R_\nu}(x) \,\mathrm{d}x = 1-\alpha.
\end{equation}
Usually $\alpha$ is set to 5\% and so the probability density function of this new RMSE emerges as a convolution of restrictions
\begin{equation}
f_{R_\nu}|_{I_{95}^\complement}(x) := \mathbb{I}_{I_{95}^\complement}(x) \cdot f_{R_\nu}(x)
\end{equation}
where $\mathbb{I}$ is the indicator function. 
Similarly to the classic RMSE we refer to this more natural metric as the significant RMSE (\textbf{sRMSE}). The sRMSE guarantees a comparison between different systems with much lower probabilities of error, simply by not taking into account the stabilising centre of all the feedback distributions. As the RMSE amplifies the remaining extreme values by its quadratic term (see Equation \ref{eq:RMSE}), resulting distributions rapidly differ under increase of false predictions.
Using this algorithm, we computed the error probabilities on a test record constructed as above. The results are depicted in Figure \ref{fig:ErrSRMSE}.
It is obvious that the error curve drops significantly faster for the sRMSE than for the RMSE and thus falls below the critical limit of 5\% much earlier. This proves that slight modification of existing quality metrics (i.e. sensitising them for human uncertainty) result into much better distinguishability of different systems as well as lower probabilities of error for a ranking.


\section{Discussion and Conclusion}

The core of this contribution is to prove the impact of human uncertainty on the comparative assessment of data mining approaches, to further specify this impact in a qualitative and quantitative manner, and finally to find solutions to the problems raised. The key messages to be shared with the data mining community are as follows:\\[.25ex]

\textsc{User Feedback are Distributions.}
Based on latest research in the field of neuroscience, action-coordinating cognitions are based on perceptions in the form of distributions which are constantly updated by a complicated generative process within the human cortex. In consequence, results of human decision making yield a certain degree of volatility and must be seen as a distribution itself. This volatility - which we denote human uncertainty in our context- can be explained by the irregular release of neuromodulators like dopamine and acetylcholine. This volatility of user feedback has been independently discovered in a simple user study.\\[.25ex]

\textsc{Metrics of Distributions become Distributions themselves.}
Based on latest research in metrology (science of accurate measurement), the uncertainty of quantities propagates with respect to a specific mathematical model when composed quantities are computed. In a probabilistic sense, the composed quantity is distributed by a probability density which emerges as a convolution of all arguments' densities. Typical approaches of determining the resulting distribution are Monte-Carlo-Simulation as well as the Gaussian Error Propagation.\\[.25ex]

\textsc{Every Ranking is Subject to an Error Probability.}
Transferred to the comparative assessment of data mining approaches, the results of our well-established precision-quality metrics turn out to be distributions rather than single scores. It is not unusual that two such distributions have an intersection, inducing a probability of error when building a ranking. This error can be thought like this:  Although a ranking according to the expected values may imply System 1 to be better than System 2, it does (more or less frequently)  occur that System 2 even outperforms System 1 when considering only single draws from the underlying distributions (representing repeated rating trials). The frequency for this ranking inversion can be seen as a probability of error that is associated to each ranking. When this error probability is too high, there is no sufficient evidence for any possible ranking and both systems become undistinguishable by means of a ranking.\\[.25ex]

\textsc{We Suggest Possible Solutions.}
After a profound elaboration of the nature of uncertainty propagation and distinguishability, we observed that big data is a fair, but by no means a complete solution, i.e. additional data (as well as the monetary expense that comes along with it) quickly loses its mending abilities. We proposed a different solution strategy which, put simply, only considers deviations that can be associated to the system's quality rather than being in range of human uncertainty. This approach is a pre-processing step and can therefore be adapted to any of the established precision quality metrics. Together with big data, this slight modification of the existing status quo should substantially improve our comparison techniques, making it even more successful.\\[.25ex]

In conclusion, our results reveal a human property that has a great impact on our choices in comparative assessments in predictive data mining. They hence justify an even more differentiated consideration of metric outcomes and mandatory pre-processing steps can be implemented with ease. Nevertheless, the research on human uncertainty within data mining approaches is not yet finished and there is still a lot more to discover. For example, it may be fruitful to further include findings from neuroscience and behavioural decision making into data mining research.
We hope that this contribution will affect its reader and lay the foundation for these possible improvements on evaluations within the field of applied data mining. To this end, all data records and algorithms used in this contribution are available open access for reproducibility and promotion of further research:  \texttt{https://jasbergk.wixsite.com/research}.




%
%
%

\bibliographystyle{IEEEtran}
\begin{scriptsize}
\bibliography{Literature}
\end{scriptsize}

%
%
%
%
%
%

\end{document}